Steven Meyer                                                                            March 2010

# Breaking GSM with rainbow Tables

## Abstract

Since 1998 the GSM security has been academically broken but no real attack has ever been done until in 2008 when two engineers of Pico Computing (FPGA manufacture) revealed that they could break the GSM encryption in 30 seconds with 200'000$ hardware and precomputed rainbow tables. Since then the hardware was either available for rich people only or was confiscated by government agencies. So Chris Paget and Karsten Nohl decided to react and do the same thing but in a distributed open source form (on torrent). This way everybody could "enjoy" breaking GSM security and operators will be forced to upgrade the GSM protocol that is being used by more than 4 billion users and that is more than 20 years old.

## GSM Security

When an operator signs a contract with a client, he gives the client a SIM card that contains firstly the IMSI (International Mobile Subscriber Identity) which is a unique 15 digit number that indicates the country, operator and mobile number and secondly a secret 128 bit key $K_i$ that is used for authentication and encryption. With these two elements, the operator pretends to guarantee Authentication (unidirectional) and privacy (that will be proven broken) of cell phone users.

When a cell phone is connecting to a network there is a phase of authentication (the bill has to be sent to the right person). The phone first sends his IMSI to the network; the network then forwards it to the home operator, if they are different (for example while traveling abroad). The home operator creates, with the secret $K_i$, several triplets containing a challenge that can be solved with the secret key, the response to the challenge and a session key $K_c$ . The network then forwards the challenge to the cell phone where the SIM card does the computation (since the secret key $K_i$ can't leave the SIM) and generates from the challenge the session key $K_c$ and the response. The cell phone then forwards the response to the network and if it is correct the operator will start to communicate with the cell phone using encrypted message with the session key $K_i$. One of the first messages sent to the cell phone will be the TMSI (Temporary Mobile Subscriber Identity) which is a random number that will be used to replace the IMSI during the next authentication with this network (to avoid the possibility of tracking a device by listening to the IMSI sent). More info on slide 8

The GSM Association offers 3 different (in)security levels, which were kept as a trade secrets that only mobile phone manufactures and mobile phone operators could buy until it was reverse engineered and made public. The original is A5/1 uses a 64 bit key and is mainly used in Europe. The A5/2 is a less secure version of the A5/1 that was invented because of local restriction in some countries and has easily been broken for a long time. Finally the A5/0 that simply does not encrypt anything is broadly used in some countries (like India) and by mobile operators when they have some technical problems or are overloaded. In theory every mobile phone should indicate if the connection is secure or not by showing a message on the screen, but operators have the ability to deactivate the warning and always do so.

## Man in the middle

One of the oldest GSM attack is a simple man in the middle attack. Cell phones have a list of authorized networks to which there are allowed to connect and they simply connect to the strongest one available. The attack consists in spoofing the operator MCC/MNC (mobile country code and network code) with a fake Base Station that has a very strong signal. When the victim tries to make a phone call the fake base station will make a connection with A5/0 with the victim and do a second normal connection with the real operator (to close the link). The original name of this attack is IMSI catcher because it was possible to ask for the IMSI instead of the TIMSI and therefore easily track someone. The only down point of this attack is that it can only work on one cell phone at the time since it has to create a connection with the real base station.

## Rainbow Table

The cryptographic function in A5/1 uses 3 linear feedback shift registers with irregular clocking. The registers are filled up with the session key (19 bits for the first one, 22 bits for the second and 23 for the third). A registry advances when his shift bit (in the yellow) is equal to the shift bit of one of the two other registries (the false assumption was that it would not be possible to backtrack a previous state). A5/2 then encrypts every "burst"(114 bits) with the xored value of the left bit on every 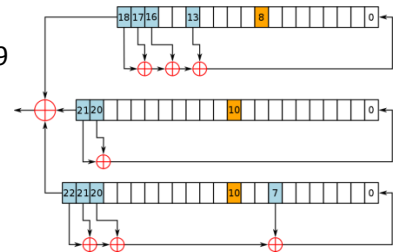 registery. The vunlnerability is that some (actualy 204 but new one are coming up all the time) known plain texts are sent encrypted which means that there are xored with the output of the registers, and therefor buy sniffing we can find the output of the A5/1 algorithm.

Now the problem is simply doing a match between A5/1 output and the internal state of the registries. A solution could be to do a dictionnary that contains all the different states of the registery and map them to A5/1 output but that would generate very big file ($10^{21} exaByte$).

Since we can't create a dictionnary attack on a 64 bit key the idea is to use "Time Memory Trade-Off" with a rainbow table.

To understand Rainbow tables it is important to understand precomputed tables . The idea is to do save only a partial part of the table in the precomputing state and then recompute the missing part during the attack. For this we need the function that we want to break (A5/1) and reduction function (like doing a xor on the entry with a constant). We strat a chain (line in the table) with a random value, calculate m times the A5/1 function 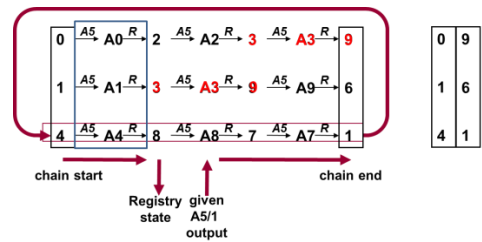 then the reduction of it, and output the result. We'll only keep the first and last value of the chain, when the other values are going to be needed there will be recalculated. This operation has to be done n times to have n lines in our table.

while creating the table we might find twice the same value at a certain point, by concequence all the next values are going to be exactly the same. This is called a collision, and to avoid having to many of duplicated data the lignes are merged togather (we ignore the newest line).
During the attack When we want to find the inverse of the A5/1 function (to find the registries state) we claculate servel times the reduction function and A5/1 on the sniffed packet (without the plaintext) until we find a value that is in the last colomn. Then we know on whitch line to search for the pre-image. So we start again the A5/1 and reduction on the first value of our line and go forward until we find the state prior our initial input and that will be what we are looking for.

While doing the attack at every step we must check if our value appears in the last column (so we can find on which line we have to be looking for). Since a lookups mean a disk access they are very expensive in time. To optimize the number of disk access we thus impose that the last colomn must have the last 12 bits set to 0. With this there are only lookups on the disk if the actual value has the last 12 bits at 0, but the risk of collision is high and that means more merges. Usualy the bigger is the table, the bigger is the collision probability, so there is always a limit in the size of a table.

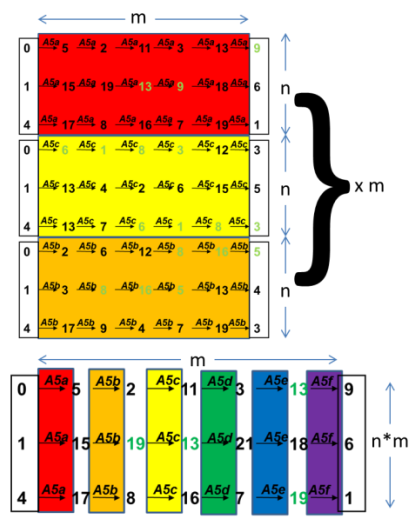

Collisions create false positives, since we have merges when two lines collide, while doing the search we might find a good end value but it won't be a first value that leads to our result (this is because we are using non-injective functions) and therefore we will never find the value we are looking for.
To avoid collisions we therefore create a new table with an other reduction function, thus, if for a certain value there is a collision, the next calculated value will be different and we won't need to merge them together. This solution brings another problem: the more tables we have, the more calculation we need to do. This means that for m tables of a length of m we need to do $m^2$ operations (and we are limited in the height because of collisions).
The rainbow table idea was that instead of creating new tables with different reduction function, we create merged tables and consider each table as a column. Now if we find two times the same value but not in the same column it won't create a collision because at the next step the output will be different since the reduction function is different. To have a real collision we need to have twice the same value in the same column and since this is quite rare, the tables can be much bigger. The research in the table is quite different. Since we don't know what reduction to start with to find the final value, we just start with the last reduction and then go backwards by testing the second last reduction followed by the last reduction and so on and so forth. This will generate in the worse case scenario 1+2+3+4…+m operations that are equal to $\frac{m*(m+1)}{2}$ operations which is more a less half the time needed with the multiple tables method. There is obviously a fifty percent chance that the result will be found in the first half of the chain's lookup which gives us even better results.

Now that we know how rainbow tables work let's see how it applies to our problem:
the architecture that has been used for this attack is $2^{15}$ elements per color, $2^5$ colors per chain, $2^{28.5}$ chains per table and $2^{8.5}$ tables which gives us a total of $2^{15} * 2^5 * 2^{28.5} * 2^{8.5} = 2^{57}$ keys precalculated. There are several reasons not to have done all the possible keys: firstly, the closer we come to the total number of key, the higher is that chance of having collisions and so the calculation time encreasese a lot. Secondly, more keys mean more space used for storing them and more calculation time needed to generate them. Because actually there a 204 known plain texts with different registry stats, the probability of finding the right key with the generated rainbow table is at 80% which is a sufficient for a proof of concept.

The generation has been done on GPUs (Nvidia & ATI) where they could generate up to 500 chains per second per GPU. Then every table had to be sorted by the last column value for a fast search during the attack.

When the GSM association heard about the rainbow table project against GSM they replied by saying that the main security of the GSM was not the the A5/1 but the frequency hopping. The idea is to change the frequency at a very fast rate (every 4.17 milliseconds) to avoid problems due to interference. In the military, frequency hopping is used as a encryption technique and for GSM the debate is still open: Chris Paget and Karsten Nohl are planning to attack this problem now.

## Other attacks

The A5/3, which is the newer and more secure version of A5/1, could also be attacked by our rainbow tables. The two protocols use the same key so the attacker simply listens to the challenge response in A5/3 to get the challenge, and then replays the same challenge in A5/1 with a fake base station. The cell phone will encrypt known plain text with the reused key then the attacker will find the actual states of the registers and backtrack to the original to find the secret key.

The SIM cards have a JVM installed into them in which there are a few applications (generally services offered by the operator). To simplify the user's life, operators can make unnotified updates to the software on the card so that the user can have the best experience. The problem will arise when the attacker will be able to make the transparent updates and will, for example, set the phone to make a conference call to the attacker's number every time the victim tries to call someone.

There is more and more new software coming out, offering higher security on smartphones. They use the data connection to make some encrypted voice-over-IP communication. The software is generally very expensive and some people are already trying to break them to (http://infosecurityguard.com/?p=26).

The point of the project is not to cause global panic and make the users go back to the Stone Age, but to show publicly that GSM is not secure and that people with bad intentions (and money) can abuse it.